# Analysis of taste heterogeneity in commuters' travel decisions using joint parking– and mode–choice model: A case from urban India

Janak Parmar, Gulnazbanu Saiyad, Sanjaykumar Dave

*Department of Civil Engineering, Faculty of Technology and Engineering, The Maharaja Sayajirao University of Baroda, Vadodara, India*

## Abstract

In developing countries like India, the policymakers have largely focused on supply-side measures, yet demand-side measures have remained unaddressed in policy implications. Ample literature is available presenting responses of TDM strategies, but most studies account mode choice and parking choice behaviour separately rather than considering trade-offs between them. This paper seeks to fill this gap by admitting parking choice as an endogenous decision within the model of mode choice behaviour which can present more valuable insights in understanding travel behaviour. This study integrates attitudinal factors and built-environment variables in addition to parking and travel attributes for developing comprehensive estimation results. A hybrid discrete choice model with random coefficients is estimated using hierarchical Bayes approach based on the Markov Chain Monte Carlo simulation method. The results reveal significant influence of mode/parking specific attitudes on commuters' choice behaviour in addition to the built-environment factors and mode/parking related attributes. It is identified that considerable shift is occurring between parking-types in preference to switching travel mode with hypothetical changes in parking attributes. Besides, study investigates the heterogeneity in the willingness-to-pay (WTP) through follow-up regression models, which provides important insights for identifying possible sources of this heterogeneity among respondents. The study provides remarkable findings which may be beneficial to city authorities for improving TDM strategies especially in developing countries.

**Keywords:** Commuting travel behaviour; Attitudes; Built-environment; Willingness-to-pay; Parking choice; Heterogeneity; Hybrid choice model.

**Corresponding Author:** mailjanakp@gmail.com

## 1. Introduction

Transportation system is among priority areas for sustainable development (Litman, 2017) as it plays significant role in the world economy and environment. Large metropolitan cities around the world have been facing ever-increasing automobile population which is frequently cited culprit for the urban traffic congestion, health and environmental issues, and depletion in valuable lands. In India, private vehicle stock has swelled from 45.6 million in 2001 to 182.9 million in 2015 (MoRT&H, India, 2018) and this growth is expected to continue in future. As per the study by the Boston Consulting Group (2018), leading consequences like delay and vehicle idling culminate into huge economic loss, estimated at USD 22 billion per year in four major cities Delhi, Kolkata, Mumbai, and Bangalore. Sustainable transport is one that is accessible, safe, environment-friendly, and affordable in which congestion, social and economic access are of such levels that they can be sustained without causing great harm to future generations (ECMT, 2004; Richardson, 1999). Hence, the concept of transportation demand management (TDM) upholds the development of sustainable mobility through the triumph of optimally balanced transport modal share in cities. It directly impacts the TDM of each transport subsystem, including parking. In developing countries like India, the policymakers have largely focused on supply-side measures, such as increasing supply of roads and promotion of large-scale infrastructure projects like metro-rails, yet demand-side measures have not been prominently addressed in the policy debate (Chidambaram et al., 2014). For an efficient design of TDM strategy, it is important to understand travellers' needs and decision-making process.

Studying behaviour-pattern of traveller is one of the pivot points in developing policies that make a travel behaviour more sustainable (Van Wee et al., 2013). This involves discouraging a travel by private modes and encouraging public transit (PT) and non-motorized transport (NMT) use. Parking policies are playing an increasingly important role across the world as a TDM tool in order to achieve these goals. Parking charge is referred as the second-best alternative to mitigate traffic congestions (Albert and Mahalel, 2006; Kelly and Clinch, 2006), preceded only by congestion pricing, but it is more commonly used because it has better acceptance rate among the different user-groups in comparison with other restrictions (Milosavljevic and Simicevic, 2019, Chapter 7). Possible reactions to parking policies involve changes in parking type (e.g., between on- and off-street parking), parking location, transportation mode, car occupancy, and frequency and trip-time. Several past studies focused on parking as a TDM strategy (e.g., Rye et al., 2006; Simićević et al., 2013; Christiansen et al., 2017; Evangelinos et al., 2018), but most studies have considered only an aspect of private vehicle (PV) use rather than trade-offs between them. Ignoring this realm may lead to biased model estimates and impropriety in policy implications. For instance, it is likely that changes in the parking attributes viz. parking cost, search time or walking to destination elicit parking relocation (e.g., from on-street to off-street parking) instead of modal shift. In that case, a mode-shift model may decree, for example, parking price elasticity (with respect to PV choice) to be short of its actual value and subsequently may lead to wrong interpretations about the efficacy of pricing policy. Besides, parking strategies may be imposed to relieve parking demand from CBDs by enhancing access to fringe-parking and to the destination. Beholding to this set-up, it is worth investigating the parking choice as an endogenous decision within the model of mode choice behaviour.

This paper seeks to fill this gap by building a joint model of parking and mode choice and assessing the possible behavioural change that could better address the feasibility of TDM strategies. Particularly, this study develops a joint-model of travel-mode and parking-choice (on-street & off-street) behaviour for commuters with origin and destination within the boundary of Delhi-National Capital Region (NCR) in India based on revealed-preference (RP) data. The model examines the impacts of attitudes toward parking and travel-modes, parking attributes such as duration, search and walk time, and built-environment of both work and home locations on commuters' behaviour. A hybrid discrete choice model (HDCM) is estimated using hierarchical Bayes approach based on the Markov Chain Monte Carlo (MCMC) simulation method. The study also investigates the taste heterogeneity in the willingness-to-pay through regression model, which provides significant insights to identify possible sources for this preference heterogeneity among respondents. The results of such an investigation would enhance the sensitivity of the development strategies and would assist in accurate formulation of land-use modelling.

The rest of the paper is organized as follows: Section 2 reviews the literature on the interaction of travel behaviour and above-mentioned measures with additional focus on parking in brief. Section 3 presents the overview of the methodology, which contains theoretical framework of the proposed HDCM and estimation of marginal WTP. Section 4 gives a comprehensive overview of survey as well as settings and specifications of different variables analysed in the model, followed by the analysis including measurement models (section 5.1), latent variable models (section 5.2), discrete choice model (section 5.3) and WTP analysis (section 5.4), and related discussion in Section 5. Lastly, Section 6 discusses the major findings and conclusions drawn from this study.

## 2. Review of Relevant Literature

### 2.1. Work Trips and Parking

Most of the parking related research has been focused on sensitivity to parking pricing (Willson and Shoup, 1990; Fearnley and Hanssen, 2012; Nourinejad and Roorda, 2017; Parmar et al., 2021), role of parking in TDM (Gillen, 1978; Christiansen et al., 2017; Litman 2017; Mingardo et al., 2022), impacts of employer-paid parking (Aldridge et al., 2006; Su and Zhou, 2012; Pandhe and March, 2012; Brueckner and Franco, 2018). A common finding of these studies is that the parking pricing policies plays a significant role in reducing commuter's private vehicle use. For instance, Willson and Shoup (1990) analyzed the effects of parking pricing at workplaces in Los Angeles and Washington D.C. in USA, and Ottawa in Canada based on four before-and-after studies. They showed 19 to 81 percent reduced car-use by employees if the free parking charges were eliminated at workplace. Similar kind of study in Norway revealed that commuters become more positive toward a parking fee led to significant drop in car-use and control over spillover of parking to local streets (Christiansen, 2014). Su and Zhou (2012) documented that the discounts in parking charges for ride-sharers, reduction in parking supply or increased parking charges for drive-alone employees are positive steps employers could take to reduce the share of drive-alone mode of travel. However, a recent study by Mingardo et al. (2022), after analyzing different parking policies namely – pricing only, pricing and time restrictions and daily tickets only, showed that time restrictions seem more effective than pricing only strategies in managing the parking demand. Many researchers have acknowledged the time-factor as an equally strong metric apart from parking pricing which influences both travel mode- and parking location-choice behaviour (e.g., Ibeas et al., 2014; Meng et al., 2017; Yan et al., 2018) buy only a few travel behaviour studies have incorporated it to analyse its effects. Marsden (2006) noted that “less evidence is available on observed responses to excess-time, particularly the time taken between parking vehicle and the final destination of commute trips.” It may lead to a potential misspecification in mode-choice models and subsequently the policymakers have vague knowledge about an efficacy of time-factors in TDM and reducing PV use. Further discussions can be found in Young et al. (1991), Marsden (2006), Inci (2015) and Parmar et al. (2019).

In addition to reducing PV-use, relieving congestion and parking demand pressure in CBDs is one of the desirable goals of TDM. A policy based on driver’s parking location choice has potential to achieve this goal by diminishing cruising for parking (Shoup, 2006) which in turn attracted many studies to analyse driver’s behaviour towards parking type and location choice (Hunt and Teply, 1993; Hess and Polak, 2004; Ibeas et al., 2014; Chaniotakis and Pel, 2015; Meng et al., 2017; Hoang et al., 2019). Hess and Polak (2004) have developed mixed logit (ML) based parking choice model and noticed significant taste variations in time-factors such as search time for parking space and walking time to the final destination. Ibeas et al. (2014) did use stated choice data to model parking choice between on-street and underground parking locations in Santoña, Spain. They assessed the interaction terms defining the effects of the individuals’ characteristics in behaviour. Their study showed parking space search time to be more important than the time to the final destination. Hoang et al. (2019) compared multinomial logit (MNL), nested-logit (NL) and mixed-logit (ML) models to evaluate the parameters influencing motorcycle drivers’ parking location choice behaviour. A dominated ML model showed that parking pricing, walking to destination, queuing time (waiting for parking), and capacity of parking lot have significant impacts on parking choice.

Notwithstanding, most parking studies encompassed the various parking attributes apart from parking charges, utmost emphasis is given to the consequences of parking pricing policies at policy level decisions (Except few, e.g., Simićević et al., 2013, Meng et al., 2017). Additionally, a little amount of research developed conclusions on travel behaviour (i.e., mode choice) considering the abovementioned parking strategies (Litman, 2018).

### 2.2. Travel Behaviour Analysis

Researchers have identified substantial range of variables to evaluate travel behaviour, typically classified as built environment, mode-specific level-of-service (LOS) attributes, parking attributes, and subjective attitudes. The effects of parking attributes are discussed in the previous subsection. This subsection briefly discusses the literature which incorporated remaining aspects. Knowledge on how built environment shapes the travel behaviour is one of the critical elements to form sustainable strategies in land-use planning. Numerous studies have explored the potential of built environment of both residential and work locations (e.g., Frank et al., 2008; Guan and Wang; 2019) to control travel behaviour. These factors usually termed as D-factors: density, diversity, design, distance to transit, and destination accessibility which are well explained by Ewing and Cervero (2010). Additionally, they found the population and job density have smaller effects on travel behaviour. However, a few studies indicated strong correlation between travel behaviour and density (Naess, 2012; Rahul and Verma, 2017). Studies in developing countries by Zegras and Srinivasn (2006), Sanit et al. (2013) and Rahul and Verma (2017) found a positive change towards the use of NMT with improved built environment factors. Studies shown that land-use diversity has significant influence on travel behaviour. For instance, when residence, work, leisure, and entertainment locations are adjacent to each other, travel distance will be reduced and subsequently NMT based trips will be increased and car-trips will be reduced (Cervero and Radisch, 1996; Yue et al., 2016). Besides, the proximity of residence and workplace with respect to different parts of the city directly affect the urban travel. People living farther from the city center (e.g., outer fringe of the city) need to travel significantly more by motorized mode (Engebretsen and Christiansen, 2011) which reflects the poor accessibility. Also, people working near central parts have lesser car-use and most trips generally made by NMT or public transit (Rahul and Verma, 2017). Lastly, transit accessibility also plays critical role in people's travel behaviour. This may be in terms of average distance through shortest paths from the residence or workplaces to the nearest transit station, transit route density within defined area around origin/destination, distance between transit stops, or the number of stations per unit area (Ewing and Cervero, 2010).

Apart from the urban form characteristics, psychological characteristics of the individual traveller, usually known as subjective attitudes have received increased attention from the researchers in recent years. These studies demonstrated that individual behaviour pattern is rooted in psychological constructs such as values, attitudes, subjective norms, perceptions and desires (Abrahamse et al., 2009; Paulssen et al., 2013). A few studies examined the effects of travel attitudes on car-use and choice of vehicle type (Steg and Kalfs, 2000; Cao et al., 2009; Van Acker et al., 2014; Etminani-Ghasrodashti and Ardeshiri, 2015; Guan and Wang, 2019). They showed how the subjective attitudes along with land-use attributes could influence the travel behaviour. Paulssen et al. (2013) developed a mixed logit-based model to map the effects of latent variables (i.e., values and attitudes) on individual's travel behaviour. Their study acclaimed that the attitudinal factors concerning flexibility, comfort and convenience, and ownership have greater impacts on travel behaviour than more conventional mode specific LOS variables. Similarly, Steg and Kalfs (2000) posed that the beliefs, preferences, and social norms primarily determine modal choice, much more than by available alternatives. Therefore, a better understanding of people's motivations is essential in order to facilitate shift towards sustainable transport modes from private vehicles. Generally, modal responses are highly sensitive to the local conditions and competitiveness of travel modes. For example, in response to parking policy, people may shift their parking location instead of changing travel mode. As per Marsden (2006), shift in parking locations is more probable response to parking interventions than shifting travel mode. Authors are of the opinion that this study is first-of-its-kind in India intends to explore the attitude based joint parking- and travel mode-choice behaviour simultaneously.

Willingness to pay for a unit change in an attribute is analyzed as an important part of transport economics since long. Researchers across the globe have studied WTP measures according to the context and motive of the study. Wardman et al. (2016) carried out a meta-analysis of value of time for European countries. They mentioned factors such as trip purpose, socioeconomic attributes, mode used, type of data, urban or intra-urban trip have significant influence on value of time. Drevs et al. (2013) studied effect of users' knowledge about public transport subsidies on WTP. Their study revealed that the public awareness about the public schemes in PT significantly increases the WTP. Sadhukhan et al. (2016) analyzed WTP for improvement of transfer facilities around metro station in Indian context using Random Parameter logit model. Results revealed that work trip commuters, high income group and high fare (of metro) group have high WTP for access time to metro. All in all, it is important to understand the WTP values for various attributes in order to estimate the possible additional fare which can be considered against the improvement costs. Also, as there is substantial heterogeneity in human perception and decision-making, it is important to examine the effects of users' socioeconomic characteristics on their WTP which is helpful while formulating improvement plans and policies.

## 3. Model Formulation

### 3.1. Overall HDCM Structure

The model is specified based upon the random utility theory, which assumes that an individual's choice, amongst mutually exclusive and collectively exhaustive alternatives, would be subjected to the highest utility depending on the considered parameters as well as unobserved part of this utility. As mentioned earlier, Hybrid Discrete Choice model (HDCM) framework is used to explicitly include the unobserved attitudes through latent variables (LV) in the utility functions. HDCM is an advanced type of discrete choice model with endogenous latent causal variables that improves the explanatory power as it accounts the impact of attitudes and perceptions of decision makers. An HDCM thus constitutes a set of individual structural and measurement models: 1. a structural equation describing LVs in terms of observable exogeneous variables (e.g., sociodemographic attributes in this case), 2. a group of measurement models relating LVs to the behavioural indicators (i.e., attitudinal questions), 3. a structural equation for the utilities of each alternative in choice model, given at least some utilities are affected by LVs, 4. a measurement model where observed choice serves as an indicator to explain the unobservable utility function. Figure 1 shows the visual framework of the HDCM model adopted in this study. The first two parts of the model are jointly termed as MIMIC model (multiple-indicator multiple-cause model). This whole system of equations of a HDCM can be represented as:

MIMIC model part-

$$\alpha_i = \Gamma z_i + \eta_i, \quad \eta_i \sim N(0,1) \tag{1}$$

$$I_i = \zeta_\iota \alpha_i + \varepsilon_i\,, \quad \varepsilon_i \sim MVN\left(0, \Sigma_\varepsilon\right) \tag{2}$$

where, $\alpha_i$ is endogenous latent attitudinal variable of an individual $i$, $z_i$ is a vector of exogeneous variables affecting LVs, and $\Gamma$ is vector of unknown parameters explaining the impact of exogeneous variables on LVs. $I_i$ is a vector of observable indicators of LVs, and $\zeta_\iota$ is a vector of unknown parameters which refers to the impact of LV on related indicators. $\eta_i$ and $\varepsilon_i$ are the measurement errors following standard normal distribution.

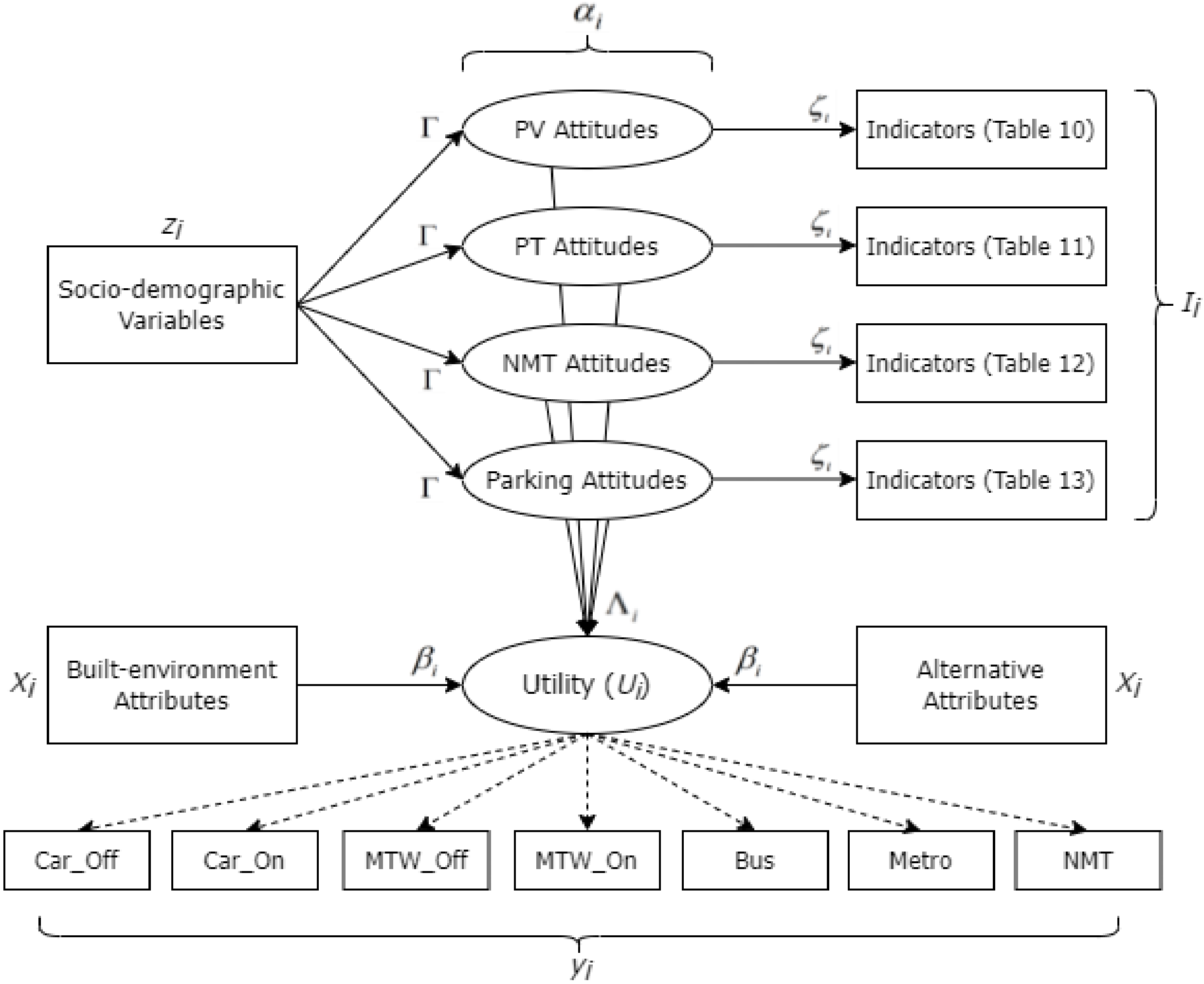

Figure 1 Conceptual framework of the adopted HDCM

Discrete choice model part-

$$U_i = \beta_i X_i + \Lambda_i \alpha_i + \in_i \tag{3}$$

$$y_{si} = \begin{cases} 1, & \text{if } U_{si} = \max_t U_{ti},\ \forall t \in C_i,\ t \neq s \\ 0, & \textit{otherwise} \end{cases} \tag{4}$$

where, $U_i$ is a vector of utilities for an individual *i*, $\beta_i$ is a vector of parameter estimates elucidating relationship of alternative attributes and built-environment factors ( $X_i$ in vector form for *i*th row) with corresponding utility, and $\Lambda_i$ is a vector of unknown parameters associated with the LVs. $\in_i$ is an error term in the utility equation and is assumed to be independent from irrelevant alternatives. It is to be noted that LVs are included in the systematic part of the utility as extra covariates to increase the explained part of the utility specification. $y_{si}$ represents the choice indicator for the respondent *i*, whose value is 1 when alternative *s* has the highest utility among the alternatives in choice set $C_i$, and 0 otherwise.

### 3.2. Ordinal Logit model for Latent variable to Indicators

The model contains latent attitudes for three predefined group of alternatives, namely PV, PT, and NMT, and these were used in the measurement model component to explain the attitudinal indicators responses. An ordered logit model was adopted to derive the likelihood of the observed responses to these indicators, which can be formulated as follows:

$$L_{I_i,q}(I_i \mid \tau,\zeta_i,\alpha_i) = \sum_{p=1}^{5} x_{I_{i,q},p} \left( \frac{e^{\tau_{I_q,p}-\zeta_{iq}\alpha_i}}{1+e^{\tau_{I_q,p}-\zeta_{iq}\alpha_i}} - \frac{e^{\tau_{I_q,p-1}-\zeta_{iq}\alpha_i}}{1+e^{\tau_{I_q,p-1}-\zeta_{iq}\alpha_i}} \right) \tag{5}$$

where $x_{I_{i,q},p} = 1$ if and only if respondent *i* selects answer *p* for the question *q*. $\zeta_i$ is parameter estimate that measures the impact of $\alpha_i$ on the attitudinal indicator $I_i$. The $\tau$ parameters are the thresholds to be estimated, with the normalization that $\tau_{I_q,0} = -\infty$ and $\tau_{I_q,5} = +\infty$. Further, as no respondents chose the lowest/second-lowest/highest level for some of the attitudinal questions, a respective second-lowest/central/second-highest level were set for $-\infty / +\infty$.

### 3.3. The Hierarchical Bayes Approach to HDCM

According to the concept, the final estimated HDCM jointly explains the logit choice probability of observing a given response and the likelihood of observing a given response of an attitudinal indictor conditional upon the latent causal variable expressed in equation (5).

The likelihood of observing the choice *s* for an individual *i* given all the unknown parameters in the utility equation (2) can be expressed as:

$$L_{si}(s \mid X_i,\beta_i,\Lambda_i,\alpha_i) = \prod \frac{e^{U_{si}}}{\sum_t e^{U_{ti}}} \tag{6}$$

It is assumed that the random term $\in_i$ is IID extreme value type 1 distributed, and thus the expression in equation (6) (also termed as discrete choice kernel) takes the Multinomial Logit (MNL) form.

Given the equations (5) and (6), the joint likelihood (non-closed form) of observing the alternative $s$ and an attitudinal indicator $I_i$ can be written as:

$$L_i\left(s,I_i|X_i,z_i,\alpha_i,\Gamma,\beta_i,\Lambda_i,\zeta_i,\tau\right)=\int_{\alpha_i} L_{si}\left(s|X_i,\beta_i,\Lambda_i,\alpha_i\right)\,L_{I_i,q}\left(I_i\,|\,\tau,\zeta_i,\alpha_i\right)\,L_{\alpha_i}\left(\alpha_i\,|\,z_i,\Gamma\right)d\alpha_i \quad (7)$$

where $L_{\alpha_i}\left(\alpha_i\,|\,z_i,\Gamma\right)$ is the density of $N\left(\Gamma z_i,1\right)$ implied to equation (1).

In the specified model, the effect of the LVs is considered as random (normal distributed in this case) across the individuals in addition to the alternative attributes, since the LVs are supposed to have a distribution, rather than fixed values and contain heterogeneity which can be reflected through predefined distribution It ensures unbiased and more certain estimators for the parameters associated with the LVs (Yanez et al., 2010). HDCM model with additional random parameters associated with the latent variables offers greater explanatory power where tastes vary with unobservable factors (i.e, latent constructs) which are purely random and heterogeneous in nature. This has been done by simultaneously integrating over the variation of LVs. Yanez et al. (2010), Tudela et al. (2011) and García-Melero et al. (2021) have posited the estimation of random coefficients associated with the LVs, though they have used a sequential approach to estimate the HDCM through classical estimation process. The classical estimation requires maximizing the loglikelihood function involving high dimensional integrals and estimated through maximum simulated likelihood estimator (Walker and Ben-Akiva, 2002). But this approach becomes very challenging and time-consuming with increased number of latent variables. Though it is possible to apply sequential approach for classical estimation, the results of this method are not efficient (Ben-Akiva et al., 2002). Therefore, this study has used a simultaneous approach to develop an HDCM model using hierarchical Bayes estimator. Readers may find more details in Bahamonde-Birke and Ortúzar (2014). The Bayesian procedure imparts two main advantages over the maximum simulated likelihood approach: (1) it avoids maximizations of any function, which can be difficult numerically, and (2) it is consistent with fixed number of draws and efficient with rise in number of draws at any rate (Train, 2009). Within a Bayesian framework, the logit models can evaluate the full posterior distribution of population-level mean vector, related covariance matrix, and the individual-specific parameters. It greatly simplifies the interrelated tasks of estimation, inference and communication compared to classical estimator (Jackman, 2000).

Consider the joint posterior distribution be $\mathcal{K}\left(.\right)$ for all the parameter required to be estimated in the choice model:

$$\mathcal{K}\left(\alpha,\theta,\Gamma,\zeta\,|\,y,I\right) \quad (8)$$

where $\theta$ contains all the parameters to be estimated in the utility equation (2). The estimates of these parameters can be obtained through different sampling methods. In this study, Gibbs sampling and Metropolis-Hastings sampling algorithms – jointly called as Markov Chain Monte Carlo (MCMC) – is used to obtain draws from the posterior distribution. At $t$th iteration, the unknowns are obtained using Gibbs samplers from the set of full conditional distributions:

$$\alpha_i^{(t)} \sim \pi\left(\alpha_i | \theta^{(t-1)}, \Gamma^{(t-1)}, \zeta^{(t-1)}, y^{(t-1)}, I^{(t-1)}\right), \forall i \quad (9)$$

$$\Gamma^{(t)} \sim \pi\left(\Gamma | \alpha^{(t-1)}, \theta^{(t-1)}, \zeta^{(t-1)}, y^{(t-1)}, I^{(t-1)}\right) \quad (10)$$

$$\zeta^{(t)} \sim \pi(\zeta \,|\, \alpha^{(t-1)}, \theta^{(t-1)}, \Gamma^{(t-1)}, y^{(t-1)}, I^{(t-1)}) \quad (11)$$

$$\theta^{(t)} \sim \pi(\theta \,|\, \alpha^{(t-1)}, \Gamma^{(t-1)}, \zeta^{(t-1)}, y^{(t-1)}, I^{(t-1)}) \quad (12)$$

For Metropolis-Hastings implementation, a draw of $\theta^{cand}$ is obtained from the distribution up to previous iteration ($t$-1) on random walk chain. The candidate realization $\theta^{cand}$ is then compared to the current $\theta^{curr}$ as:

$$\Psi = min\left\{1,\ \frac{L(\theta^{cand}|y, X)\,\pi\left(\theta^{cand}\right)}{L(\theta^{curr}|y, X)\,\pi\left(\theta^{curr}\right)} \times \frac{p\left(\theta^{cand}, \theta^{curr}\right)}{p\left(\theta^{curr}, \theta^{cand}\right)}\right\} \quad (13)$$

where $p$ ($a$, $b$) is the probability of generating candidate $b$ given $a$. $\theta^{cand}$ in the $t$th iteration is admitted as the new $\theta^{curr} = \theta^{cand}$ with probability $k$, while the old one at $t$-1 iteration is preserved as $\theta^{curr} = \theta^{curr}$ with probability 1-$k$. Then the Bayesian solution can be obtained by infusing this MH results into the Gibbs sampler over many iterations until the parameter values converge to draws from the posterior. For more details, reader may be interested to the read in Alvarez-Daziano and Bolduc (2013). The assertion of the convergence can be made by examining if the draws after burn-in period are moving around the posterior (Train, 2009).

### 3.4. Estimation of Marginal Willingness-to-pay (MWTP)

The proposed model allows random coefficients to capture the taste heterogeneity in choice behaviour. In this paper, the MWTP is analyzed for the alternative attributes whose coefficients are obtained as discussed in previous subsection. Consider the distribution of $\beta_i$ for a given sampled population be $f\left(\beta_i \,|\, \mu, \omega\right)$ with mean vector $\mu$ (normal with extremely large variance) and covariance vector $\omega$ (inverted Wishart with K degrees of freedom). In hierarchical Bayes model, the draws for $\mu$, $\omega$ and $\beta_i$ are obtained from their joint posterior defined in Equation 8. If $\overline{\beta}_{ia}$ be the coefficient of attribute $a$ and $\beta_{ic}$ be the cost coefficient for an individual $i$, then the individual-specific MWTP can be calculated as:

$$MWTP_{ia} = -\frac{\overline{\beta}_{ia}}{\beta_{ic}} \quad (14)$$

Once the individual level MWTP is estimated, the variations in calculated MWTP among the individuals can be analyzed by establishing the relationship between possible sources (e.g., socioeconomic characteristics) and MWTP through regression. This type of two-step approach was used in a very few studies in different aspects (Campbell, 2007; Hoshino, 2011) to analyze the preference heterogeneity in choice behaviour, though it is yet to be explored in the context of travel behaviour. These studies showed that this approach yielded greater validity and explanatory power than standard methods used to incorporate individual-specific variables into the analysis. The MWTP regression in this study assumes that respondents' socioeconomic characteristics primarily influence the heterogeneity in MWTP. Based on the individual MWTP calculated using Equation 14, and considering $q_{ia}$ be a vector of respondent $i$'s socioeconomic characteristics, the regression model can be expressed as:

$$MWTP_{ia} = c_a + \gamma'_a \, q_{ia} + \delta_{ia} \tag{15}$$

Where, $c_a$ is an intercept, which is constant for all individuals; $\gamma'_a$ is the coefficient vector; and $\delta_{ia}$ is an unobserved random term reflecting the effect of omitted variables.

## 4. Survey and Data Description

For this study, data were collected through face-to-face interview survey of commuters (students were not included) in Delhi, India. The urban population of the city was estimated as 26 million (97.5% share), making it the world's second largest urban areas and comprises second-highest GDP per capita in India (Planning Dept., 2019). The city has been experiencing perturbing inflation in traffic congestion and air pollution in past two decades which are the consequences of the rapidly soared motor-vehicle population. Currently, it has 11.27 million of registered vehicles in terms of cars, jeeps, and motorized two-wheelers (MTW) sharing nearly 95% of the total vehicles (Economic Survey of Delhi, 2021). Additionally, they demand huge space for parking in CBD areas to waste valuable land resources. It is estimated that the annual demand for car parking space in Delhi can be equivalent to as much as 471 football fields. Public transit in Delhi has two major services- bus (fleet size of 6572) and metro, having daily ridership of nearly 11 million as per Economic Survey of Delhi (2021). The public transit share is near to 50%, while private vehicles including cars and MTW account for about 32% of total motorized trips in the city.

This study used the data collected through RP survey conducted between September to November 2019 in Delhi. Respondents were recruited randomly at their work locations in order to complete the survey form making sure that the respondent is eligible for the interview (i.e., the trip is within study area, had an alternative transport mode and parking-type available to use). Care was taken to interview respondents arriving from all parts of the city to include all sorts of variations in the samples. A questionnaire includes the information regarding personal and household (HH) characteristics, vehicle-ownership levels, trip and parking related attributes, and access/egress details. It also comprises several modes- and parking-specific attitudinal statements on a 5-point Likert-scale from "strongly disagree" to "strongly agree". The collected data was continuously monitored while survey was in progress to ensure a good balance of data spreading across the city and having well distributed trip characteristics.

A total of 650 respondents were interviewed of which 602 samples were refined (excluding unqualified and incorrect answers; effective rate of 92.6%) to be eligible for further use in analysis. The key socio-demographic statistics of the 602 respondents is demonstrated in Table 1. It can be seen that the proportion of respondents with service as an employment status is much higher (84%) compared to other categories. In the survey data, the observations of part-time working people were very less and hence it was merged with full-time working observations to denote a merged category as a service. Also, it was noticed that near to 24% of the off-street parking users had employer-paid parking and none of their employer has parking cash-out scheme in place.

Table 1 Descriptive Statistics of analysed samples

| Individual Characteristics | Category | Frequency (%) |
|---|---|---|
| Sex | Male | 404 (67.1) |
| | Female | 198 (32.9) |
| Age | 18 – 24 | 13 (2.2) |
| | 25 – 39 | 351 (58.3) |
| | 40 – 64 | 238 (39.5) |
| Income<br>(Thousand ₹)<br>(1 INR ~ 0.014 USD) # | < 20 | 105 (17.5) |
| | 21 – 35 | 204 (33.9) |
| | 36 – 50 | 129 (21.4) |
| | 51 – 70 | 76 (12.6) |
| | > 70 | 88 (14.6) |
| Education level | Primary-school | 4 (0.7) |
| | High-school | 122 (20.3) |
| | Graduate | 393 (65.3) |
| | Postgraduate and higher | 83 (13.7) |
| Employment Status | Service | 506 (84) |
| | Business | 36 (6) |
| | Self-employed | 60 (10) |
| Current Transport Mode* | Car & Off-street parking (Car_Off) | 104 (17.3) [37.5] |
| | Car & On-street parking (Car_On) | 46 (7.6) [16.6] |
| | Motorcycle & Off-street parking (MTW_Off) | 68 (11.3) [14.1] |
| | Motorcycle & On-street parking (MTW_On) | 84 (13.9) [17.4] |
| | Public transport– Bus | 148 (24.6) [24.6] |
| | Public transport– Metro | 100 (16.6) [16.6] |
| | Non-motorized modes (NMT) | 52 (8.6) [42.3] |
| | | |
| Household Characteristics | | Mean (Std. Dev.) |
| Members in HH | | 3.83 (1.44) |
| Number of working members | | 1.65 (0.65) |
| Number of Cars in HH | | 0.53 (0.63) |
| Number of MTW in HH | | 1.06 (0.68) |
| Number of Bicycles in HH | | 0.69 (0.61) |

* The values in square brackets show *percent times chosen when available/feasible*.
[#] The conversion rate between USD ($) and INR (₹) is contemporary to the survey period.

The trip features such as trip distance, trip time (purely an in-vehicle time), trip cost, and parking features such as parking duration, search time for parking space, and walk time to destination (i.e., egress time) were considered in this study. All these indicators were self-reported by the respondents. The average travel distance across all modes was observed to be about 10 km which is also an average commute distance across the Delhi-NCR. For trip time, only in-vehicle time was considered as a trip attribute. This is because the access and egress features for public modes are taken as BE factors whereas for the private modes, access features are neglected as people park their vehicles right in-front of their homes and egress features are considered within parking attributes. Trip cost is estimated on monthly basis in this study because for PV categories, maintenance cost is also taken into account in addition to the fuel cost to demonstrate the full cost people incur for using PVs. The logic for considering full variable trip cost is that – in India, the wages are normally paid on a monthly basis unlike western countries where it is on weekly basis. The commuters estimate all household expenditures including transportation costs on monthly basis and hence it was very quick for respondents to answer questions related to travel costs. Moreover, commuters who use public transport possess monthly pass whose cost is different from regular tickets. Hence, it is quite reasonable to consider monthly costs for all the modes. To distinguish between the on-street and off-street parking, parking demand in terms of duration, search time for parking space, and egress time for both the parking types were considered. Though the parking price is one of the most important deciding factors in parking choice, it was not considered, because as per the current scenario, the parking price for both on-street and off-street parking is same in study location. Further, the current parking tariff is ₹20 per hour with the maximum of ₹100 per day for car, and ₹10 per hour with the maximum of ₹50 per day for MTW in the city. Since the minimum parking duration is greater than 5 hours as per the data, parking cost incur to PV users become constant to all individuals. In this sense, it does not have influence on commuters' parking and mode choice decision while using revealed-preference data.

### 4.1. Built Environment Factors

Using the information developed from the geo-coded land use database, several spatial characteristics of the built environment were estimated in this study. A political ward map (272 wards) was considered as a base map to identify the origin and destination of the respondents on the map. The built environment factors include land use diversity (entropy index), road density, working population density, and accessibility– distance from home and work location to the closest metro station and closest bus stop (self-reported). The density and diversity parameters were estimated based on the one-kilometre radius around the reported addresses. It should be noted that density and diversity indicators are calculated at the trip origins only.

### 4.2. Attitudinal Factors

As this study tries to identify objective as well as subjective influences on commute travel behaviour, individuals' attitudinal preferences towards travel modes and parking were captured in the survey using a 5-level Likert scale. The list of the statements regarding attitudes included in the survey is shown in the Appendix. The responses to these statements were then factor analysed using principal axis factoring method (promax rotation) to find out how they can be related to each other. The factors (latent variables) were extracted from these observed attitudes for individual modes to understand which latent variables are important to which specific mode. Cronbach's alpha was used to measure the internal consistency i.e., how closely a set of indicators for each latent variable related to each other in a same group. The obtained alpha values for each group were acceptable ($\propto > 0.7$) as listed in the respective Tables in Appendix. Additionally, Pearson's correlation matrix presented high correlation values between factors which also support the factor analysis used in this study. The Kaiser-Meyer-Olkin's (KMO) sampling adequacy test was employed to check the suitability of data for factor analysis. The latent variables obtained from the observed attitudes are: 1. Individualist, pro-environment, economy, comfort, and flexibility for PVs (68.63% variance explained, KMO = 0.73); 2. Comfort, convenience, safety, and flexibility for PT (60.76% variance explained, KMO = 0.69); 3. Health, safety, comfort, and convenience for NMT modes (71.78% variance explained, KMO = 0.81); and Safety and convenience for Parking types (59.85% variance explained, KMO = 0.71). Next, a measurement models were formulated within the HDCM framework to validate the relationships between the observed and latent variables, which is discussed in the next section.

## 5. Model Estimation and Results

### 5.1. Measurement Models

After ensuring the adequacy of the factor analysis, measurement models were specified for each transport modes group (i.e., PV, PT & NMT) as well as parking type. The estimated relationships for each indicator subjected to the respective latent variable are presented in the Tables 2, 3, 4 and 5. All the estimated coefficients had the expected sign and significant at the 90% confidence level or higher. The threshold parameters are also significant at 90% level or higher which shows that the proposed ordered models adequately reflect individual's opinions on the attitudinal statements presented to them. In each table, posterior means $\mu$ of estimates ($\zeta_{\iota}$) are listed, which are called marginal utilities of the LVs in the ordered-logit model. A negative estimate refers to the weak association i.e., respondents less likely to agree with the indicator statement as the intensity of underlying LV increases. The respective thresholds show the required change in the utility to make the association stronger/weaker between LV and indicators.

Table 2 Measurement equations in the MIMIC model for PVs

| Indicator | Estimate | Thresholds | | | |
|---|---|---|---|---|---|
| | | 1 | 2 | 3 | 4 |
| Individualist1 | 1.917 | -4.817 | 1.304 | 4.750 | |
| Individualist2 | 0.836 | -3.205 | -1.661 | 1.225 | 2.669 |

| Individualist3 | 0.167 | -2.351 | -0.274 | 0.720 | 2.698 |
|---|---|---|---|---|---|
| Individualist4 | 0.094 | -1.325 | -0.190 | 1.445 | |
| Pro-environment1 | 0.661 | -5.777 | -1.181 | 1.800 | 5.732 |
| Pro-environment2 | 0.721 | -1.601 | 0.502 | 3.109 | 7.852 |
| Pro-environment3 | 0.185 | -0.778 | 0.428 | 2.321 | |
| Economy1 | 1.160 | -5.532 | -0.171 | 1.846 | 6.185 |
| Economy2 | 2.061 | -7.599 | -1.099 | 0.107 | 4.053 |
| Economy3 | 0.340 | -2.656 | -0.777 | 0.097 | 2.183 |
| Comfort1 | -0.386 | -3.700 | -1.907 | -1.067 | 1.026 |
| Comfort2 | 1.114 | -3.764 | -1.140 | 0.794 | 4.438 |
| Flexibility1 | 0.818 | -0.008 | 2.849 | | |
| Flexibility2 | -0.783 | -3.363 | 0.030 | 1.723 | 4.179 |
| Flexibility3 | -0.596 | -3.791 | -1.549 | 0.359 | 4.291 |

Table 3 Measurement equations in the MIMIC model for PT

| Indicator | Estimate | Thresholds | | | |
|---|---|---|---|---|---|
| | | 1 | 2 | 3 | 4 |
| Comfort1 | 0.458 | -1.937 | -0.596 | 0.833 | 2.843 |
| Comfort2 | 0.641 | -3.344 | -0.400 | 1.545 | 4.537 |
| Comfort3 | 0.815 | -3.284 | -0.183 | 2.510 | |
| Comfort4 | 0.309 | -4.746 | -1.744 | 0.224 | 2.341 |
| Convenience1 | -0.243 | -5.546 | -3.108 | -1.458 | 0.626 |
| Convenience2 | -0.124 | -1.005 | 0.301 | 1.210 | 3.561 |
| Convenience3 | 0.296 | -3.960 | -3.202 | -1.561 | 1.484 |
| Safety1 | -0.321 | -2.932 | -1.677 | -0.458 | 1.136 |
| Safety2 | 0.296 | -4.535 | -2.090 | 0.339 | 2.432 |
| Safety3 | 1.289 | -1.407 | 0.143 | 1.833 | 4.522 |
| Flexibility1 | 0.489 | -5.125 | -2.447 | -0.900 | 1.735 |
| Flexibility2 | 0.966 | -4.716 | -2.018 | -0.679 | 1.601 |
| Flexibility3 | 0.429 | -5.482 | -2.439 | -0.715 | 2.144 |

Table 4 Measurement equations in the MIMIC model for NMT

| Indicator | Estimate | Thresholds | | | |
|---|---|---|---|---|---|
| | | 1 | 2 | 3 | 4 |
| Health1 | 1.563 | -4.209 | -1.221 | 2.038 | |
| Health2 | 0.189 | -4.486 | -1.109 | 0.416 | 1.790 |
| Health3 | 1.295 | -4.106 | -1.900 | 0.514 | 4.008 |
| Safety1 | 0.080 | -2.327 | 0.314 | | |
| Safety2 | 0.815 | -2.285 | 0.938 | 2.809 | |
| Safety3 | 2.382 | -4.483 | 1.610 | 5.262 | |

| Comfort1 | 1.134 | -6.803 | -2.345 | 1.920 | |
|---|---|---|---|---|---|
| Comfort2 | 0.255 | -3.326 | -0.618 | 1.410 | |
| Convenience1 | 0.144 | -2.841 | -1.227 | 0.824 | |
| Convenience2 | 0.067 | -4.425 | -0.003 | | |

Table 5 Measurement equations in the MIMIC model for parking types

| Indicator | Estimate | Thresholds | | | |
|---|---|---|---|---|---|
| | | 1 | 2 | 3 | 4 |
| Safety1 | 0.989 | -3.751 | -1.504 | 0.114 | 2.289 |
| Safety2 | 1.507 | -2.497 | -0.131 | 2.563 | |
| Convenience1 | 3.334 | -3.718 | 0.200 | 2.226 | 6.416 |
| Convenience2 | -0.073 | -2.597 | -0.611 | 1.873 | |
| Convenience3 | -2.218 | -5.494 | -1.996 | -0.050 | 2.949 |

Looking to the results of the measurement models, some of the LVs were redefined in the model as the overall relationship between indicators and underlying LV confirms the negative association between them. For example, the *safety* in PT is redefined as *less safe* because overall effects of three indicators influence negatively on PT choice. Similar are the cases for comfort in PT, safety and comfort in NMT. For parking types, safety should be taken as positive for off-street parking and negative for on-street parking, and vice versa for convenience. These LVs are written with negative "-" sign in further discussion. Reader should refer tables in Appendix for more clarification.

### 5.2. Structural Models for Latent Variables

The second part of the MIMIC model consist of a structural model seeking to understand how travellers' observed characteristics affect their latent attitudes. Initially, the sociodemographic and travel characteristics were included to examine their effects on LVs, however only the former were found significant and hence shown here. The results are as expected and analogue to the previous findings. The estimates of these models are to be interpreted in conjunction with the related measurement models tabulated in section 5.1.

Table 6 presents the results for the structural model of five latent variables specified for commuting by private vehicles. Females, graduates, and respondents with higher income (above 50k) tend to be more individualist. Respondents aged under 25 and in 25-40 group have less positive value for individualist attitude compared to the base age group of above 40. Female respondents tend to be more pro-environment compared to males, as are the graduates. Individuals in age group of 25-40 are more positive on environmental concerns compared to younger and older aged respondents. Females seem in less agreement with the economic indicators to restrict the PV usage compared to males, which is similar to one observed for the higher income group respondents. Also, it is observed that the respondents with higher number of working members in family have similar attitude towards economic indicators. The coefficient of comfort for gender is less, indicating no major gender-wise distinction towards this attitude. Aged people are more concerned for comfort, as are the people with higher income and graduates. Males put more value for flexibility compared to females and similar is the case for higher income group respondents. Respondents aged below 40 have more importance for flexibility than those aged above 40.

Table 6 Structural equations model for latent variables (PV)

| Effect from… ↓<br>On… → | Individualist | Pro-environment | Economy (-) | Comfort | Flexibility |
|---|---|---|---|---|---|
| Female (vs male) | 0.641 | 0.445 | -0.564 | 0.030 | -0.741 |
| Age under 25<br>(vs age above 40) | -0.386 | -2.192 | 1.342 | -1.509 | 0.830 |
| Age 25-40<br>(vs age above 40) | -0.226 | 0.951 | 0.269 | -0.559 | 0.475 |
| Graduate<br>(vs non-graduate) | 0.597 | 0.674 | 0.282 | 0.286 | -0.391 |
| Income under 20k<br>(vs 35k-50k) | -0.781 | - | 0.225 | -1.808 | -0.121 |
| Income 20k-35k<br>(vs 35k-50k) | 0.080 | - | 0.523 | -0.449 | -0.414 |
| Income 50k-70k<br>(vs 35k-50k) | 0.522 | - | -0.533 | 0.235 | 0.383 |
| Income above 70k<br>(vs 35k-50k) | 1.257 | - | -0.952 | 0.660 | 0.404 |
| No. of working members | - | - | -0.268 | - | - |

Table 7 Structural equations model for latent variables (PT)

| Effect from… ↓<br>On… → | Comfort (-) | Convenience | Safety (-) | Flexibility |
|---|---|---|---|---|
| Female (vs male) | 0.678 | -1.107 | 0.618 | -1.035 |
| Age under 25<br>(vs age above 40) | -0.932 | 1.979 | 0.193 | -0.063 |
| Age 25-40<br>(vs age above 40) | -0.189 | -0.276 | 0.381 | 0.748 |
| Graduate<br>(vs non-graduate) | 0.660 | -0.078 | -0.225 | -0.027 |
| Income under 20k (vs 35k-50k) | -1.507 | 4.361 | -0.659 | 0.791 |
| Income 20k-35k (vs 35k-50k) | -0.475 | 1.420 | -0.466 | 0.821 |
| Income 50k-70k (vs 35k-50k) | 0.832 | -1.429 | 0.649 | -0.012 |
| Income above 70k (vs 35k-50k) | 1.752 | -1.668 | 0.967 | -0.872 |

Table 7 shows underlying structural relationship between sociodemographic characteristics and LVs related to the use of PT modes. In the column of comfort, it is seen that females and graduates are more concerned on comfort in PT modes, as is the case for the respondents with increased income and aged above 40. Looking to convenience, males find PT modes less convenient compared to females, as are the respondents with age under 25 and above 40. The same applies to the non-graduate respondents, though the intensity of coefficient is not much higher. Comparing the individuals from various income groups, it can be seen that lower income individuals are more concerned about convenience compared to individuals with higher income. The results for concern for safety indicate that the women, younger people, non-graduates, and higher income people are more concerned about safety if they use PT modes. Lastly, it can be observed that females feel PT modes less flexible, as are non-graduates and higher income respondents. Compared to individuals aged above 40, the individuals in the age group of 25-40 find PT more flexible which contrasts with the respondents aged below 25.

Table 8 Structural equations model for latent variables (NMT)

| Effect from… ↓<br>On… → | Health | Comfort (-) | Convenience | Safety (-) |
|---|---|---|---|---|
| Female (vs male) | -0.205 | 0.361 | -1.710 | 1.578 |
| Age under 25 (vs age above 40) | 0.693 | 1.536 | -1.235 | -0.538 |
| Age 25-40 (vs age above 40) | -0.099 | -1.187 | 1.996 | -0.287 |
| Graduate (vs non-graduate) | 0.382 | 0.534 | -0.981 | 0.615 |
| Income under 20k (vs 35k-50k) | - | - | 3.681 | -0.934 |
| Income 20k-35k (vs 35k-50k) | - | - | 1.950 | -1.015 |
| Income 50k-70k (vs 35k-50k) | - | - | -0.451 | 1.413 |
| Income above 70k (vs 35k-50k) | - | - | -3.398 | 1.616 |

The results in Table 8 show that males are more positive on health benefits of using NMT modes, similar to graduate respondents. Also, it is noticed that younger and older respondents more health benefits of NMT use compared to respondents aged between 25-40. Looking to the concerns regarding comfort, females seek more comfort while commuting through NMT modes, as are the graduates. It is found that respondents aged between 25-40 are less concerned on comfort compared to other two categories. With respect to the convenience, men consider NMT modes significantly more convenient compared to women for shorter trips, as are the respondents aged between 25-40 compared to others. It is notable that lesser income group respondents perceive NMT more convenient, and it deteriorates as the income increases. From the safety point of view, females are observed to be more concerned about safety compared to males and so are the respondents aged above 40 compared to those aged below 40. Graduates are more apprehensive about NMT safety compared to non-graduates, which is in contrast with what is observed for PT safety. While observations for the effects of income are like those found for PT.

Table 9 Structural equations model for latent variables (off-street vs. on-street parking)

| Effect from… ↓<br>On… → | Convenience (-) | Safety |
|---|---|---|
| Female (vs male) | -0.059 | -0.123 |
| Age under 25<br>(vs age above 40) | -0.432 | 0.815 |
| Age 25-40<br>(vs age above 40) | -0.303 | 0.186 |
| Graduate<br>(vs non-graduate) | -0.326 | 0.830 |

The coefficients in Table 9 shows that females are more concerned for both the LVs compared to males while choosing parking type. Looking to the age, it is observed that the respondents above 40 are more concerned about convenience which is in a contrary to what observed for the LV related to safety. The results also indicate that non-graduate respondents are more concerned about convenience while graduate ones are more concerned about safety.

### 5.3. The Discrete Choice Model

In what follows, the latent variables were introduced into the discrete choice part of HDCM together with the other parameters discussed previously. Several trial-and-error iterations were carried out by adding and removing different variables considering their sign, significance, and importance in the model in order to obtain the best specifications for the model. To estimate the proposed model, MCMC algorithm was performed with 200,000 draws of which 100,000 sweeps were discarded as burn-in. The next 100,000 sweeps were retained to estimate the posterior means of considered parameters. At the starting of the MCMC run, informative priors were defined for these parameters taking into account previous studies cited in literature review section as well as pre-developed MNL models.

Table 10 shows the choice model component of HDCM. A variable choice set including Car_Off, Car_On, MTW_Off, MTW_On, Bus, Metro and NMT alternatives was considered in the model estimation. The model explains the effects of LVs, built-environment factors, and level-of-service variables on the choice decision. The results are broadly conformed to the expectations. The final loglikelihood value depicts significant improvement over null model as it can impart sufficient information on individuals' choice behaviour. The underlying mean and standard deviation estimate for the random effects based on posterior normal distribution are reported. The coefficients for entropy index and both the density parameters are kept fixed and hence only the mean value is reported. A significant heterogeneity in more or less amount is observed for all parameters, as evidenced by the significant standard deviations. A few non-significant parameters were also retained in model as they provide important information regarding respondents' decision-making.

The ASCs in the model show that all the modes but metro was preferable over bus given the availability. Looking to the model, almost all considered latent variables are significantly related to respondents' choice behaviour, indicating that these parameters are determinants for mode as well as parking choice. PV_Individualist attitude has a significant positive impact on private mode use for commuters. The proposed model allows to calculate the probability given the coefficient for an attribute as the density function for the coefficient vector $\beta_i$, $N(\mu,\omega)$ is now known. Hence, the probability given the positive coefficient for PV_Individualist for car can be estimated as P($\beta_{PV_Individualist}$ > 0) = 0.999, which suggests that almost all respondents prefer to use car when their attitude is positively associated with the related response to the statements. Similar is the case for MTW mode (P = 0.947). In contrary, the awareness on environmental concerns causes negative effects on both the private modes, but the magnitude of coefficient is higher for car. It dictates that the use of car is substantially dependent on the pro-environment attitude (P($\beta_{PV_Pro-environment}$ < 0) = 0.998). It can be noted that commuting by MTW is positively related to the attitude of economy. It reveals that respondents are less concerned about economic aspects which may be because the current levels of travel cost and parking fees are comparatively much lesser for MTW. The attitude related to travel comfort is positively associated with the choice of private modes and supports the statements for the same. The effect is stronger for car compared to MTW looking to the magnitude of coefficients. Flexibility has overall negative effect on the car use while it is positive for MTW. It reveals that respondents feel MTW more flexible compared to car. The model also suggests that car users put more weightage on safety while selecting parking type. Conversely, MTW users consider convenience as more important compared to safety. The related probabilities are estimated as: P($\beta_{Parkng_Safety}$ > 0) = 0.991 for Car_Off; P($\beta_{Parkng_Safety}$ > 0) = 0.908 for MTW_Off; P($\beta_{Parkng_Convenience}$ > 0) = 0.747 for Car_On; and P($\beta_{Parkng_Convenience}$ > 0) = 0.999 for MTW_On. These state that the parking users with safety concerns in their mind tend to park their vehicle at off-street parking, whereas individuals would prefer the on-street parking given their positive attitudes towards parking convenience.

1 Table 10 Estimated parameters for the choice model (t-stats in the parenthesis)

| Variable | Car_Off | | Car_On | | MTW_Off | | MTW_On | | Bus | | Metro | | NMT | |
|---|---|---|---|---|---|---|---|---|---|---|---|---|---|---|
| | Mean | SD | Mean | SD | Mean | SD | Mean | SD | Mean | SD | Mean | SD | Mean | SD |
| ASCs[#] | 2.088 (8.70) | 1.140 (3.21) | 2.470 (2.11) | 1.854 (3.65) | 4.129 (11.86) | 2.914 (6.82) | 9.381 (6.44) | 2.484 (1.85) | - | - | -1.899 (-2.04) | 1.560 (3.57) | 2.684 (1.66) | 1.691 (0.18) |
| **Latent variables** | | | | | | | | | | | | | | |
| PV_Individualist[*] | 1.389 (2.43) | 0.208 (0.82) | | | 0.547 (1.833) | 0.337 (1.77) | | | - | - | - | - | - | - |
| PV_Pro-environment[*] | -1.720 (-6.09) | 0.566 (7.72) | | | -0.498 (-2.85) | 0.432 (1.75) | | | - | - | - | - | - | - |
| PV_Economy[*] | -1.507 (-2.30) | 0.504 (2.18) | | | 1.595 (2.05) | 0.733 (5.27) | | | - | - | - | - | - | - |
| PV_Comfort[*] | 1.669 (1.68) | 0.401 (1.73) | | | 0.176 (2.54) | 0.819 (0.37) | | | - | - | - | - | - | - |
| PV_Flexibility[*] | -0.098 (-4.95) | 0.008 (2.31) | | | 0.701 (5.27) | 0.017 (2.47) | | | - | - | - | - | - | - |
| Parking_Safety | 2.127 (2.40) | 0.892 (1.95) | -2.087 (-2.34) | 0.341 (2.09) | 1.072 (1.70) | 0.807 (2.02) | -1.106 (-2.93) | 0.537 (2.14) | - | - | - | - | - | - |
| Parking_Convenience | -1.018 (-2.89) | 0.514 (2.11) | 0.521 (2.80) | 0.782 (1.16) | -1.179 (-3.07) | 0.354 (1.69) | 3.179 (2.94) | 0.387 (2.85) | - | - | - | - | - | - |
| PT_Comfort[*] | 1.713 (1.06) | 0.295 (2.23) | | | 2.790 (4.90) | 1.051 (2.10) | | | 0.693 (4.02) | 0.456 (2.28) | -0.957 (-0.74) | 0.280 (1.94) | -0.497 (-2.30) | 0.444 (2.01) |
| PT_Convenience[*] | 0.634 (2.47) | 0.098 (1.85) | | | 1.740 (2.17) | 0.867 (1.72) | | | -1.624 (-2.80) | 0.767 (4.66) | 2.456 (0.69) | 0.489 (3.07) | -2.241 (-4.18) | 0.838 (4.34) |
| PT_Safety[*] | 2.026 (5.98) | 0.588 (1.76) | | | 3.498 (6.31) | 0.450 (0.25) | | | -2.924 (-4.68) | 1.294 (1.78) | -1.732 (-2.32) | 0.364 (2.21) | 0.747 (2.33) | 0.422 (0.92) |
| PT_Flexibility[*] | -1.829 (-1.54) | 0.722 (4.34) | | | 1.550 (2.17) | 0.254 (2.09) | | | 1.319 (1.94) | 0.873 (0.14) | 1.597 (5.44) | 0.704 (1.86) | -0.791 (-1.76) | 0.347 (2.59) |
| NMT_Health[*] | -1.513 (-1.78) | 0.262 (2.13) | | | -1.932 (-2.12) | 0.242 (2.07) | | | -2.498 (-1.96) | 0.265 (2.48) | -1.321 (-4.21) | 0.591 (1.42) | 3.539 (2.36) | 0.330 (6.07) |
| NMT_Safety[*] | 1.163 (6.66) | 0.515 (4.30) | | | 1.345 (2.07) | 0.347 (5.52) | | | 1.143 (3.38) | 0.325 (1.86) | 0.261 (4.20) | 0.423 (2.04) | -1.957 (-7.54) | 0.450 (2.33) |
| NMT_Comfort[*] | 1.251 (3.86) | 0.176 (2.47) | | | 1.359 (1.91) | 0.364 (4.06) | | | -1.982 (-1.88) | 0.367 (1.68) | 0.600 (3.82) | 0.207 (1.89) | -0.988 (-4.33) | 0.492 (1.75) |

| NMT_Convenience* | -1.629 (-2.43) | 0.354 (1.75) | | | 0.564 (3.88) | 0.942 (1.97) | | | 2.274 (2.60) | 0.408 (4.10) | -3.205 (-2.04) | 0.333 (1.91) | 2.389 (4.58) | 0.692 (1.83) |
|---|---|---|---|---|---|---|---|---|---|---|---|---|---|---|
| **Built-environment parameters** | | | | | | | | | | | | | | |
| Access distance– Bus* | 0.711 (3.31) | 0.245 (1.89) | | | 1.284 (2.30) | 0.568 (0.97) | | | -2.263 (-1.93) | 0.590 (2.40) | 0.939 (2.52) | 0.282 (2.22) | - | - |
| Access distance– Metro* | 1.106 (3.83) | 0.219 (2.30) | | | 1.210 (2.40) | 0.415 (2.16) | | | 0.982 (2.01) | 0.233 (1.96) | -2.360 (-1.87) | 0.202 (2.49) | - | - |
| Egress distance– Bus* | 0.362 (1.67) | 1.004 (1.38) | | | 1.118 (1.97) | 0.324 (0.21) | | | -3.406 (-4.57) | 0.441 (5.79) | 0.524 (2.16) | 0.605 (1.70) | - | - |
| Egress distance– Metro* | 0.517 (1.72) | 0.272 (6.57) | | | 0.709 (5.13) | 0.298 (2.83) | | | 0.181 (4.02) | 0.278 (3.31) | -2.779 (-1.96) | 0.365 (2.01) | - | - |
| Entropy index* | -3.831 (-1.67) | | | | -2.413 (-4.96) | | | | 2.654 (1.97) | | 0.966 (4.18) | | -0.587 (-2.49) | |
| Residential density* | -1.515 (-4.47) | | | | -0.408 (-2.24) | | | | 1.643 (4.39) | | 1.161 (2.15) | | 1.259 (1.80) | |
| Working population Density* | 0.495 (0.83) | | | | 1.405 (4.51) | | | | 1.266 (5.14) | | 0.703 (3.65) | | 3.978 (1.84) | |
| **Parking and Travel parameters** | | | | | | | | | | | | | | |
| Employer paid parking | 1.221 (3.26) | 0.564 (2.90) | - | - | 2.518 (2.49) | 0.861 (4.65) | - | - | -0.931 (-0.51) | 0.274 (1.13) | -0.397 (-1.77) | 0.069 (1.88) | - | - |
| Parking duration | 2.586 (1.88) | 0.255 (2.31) | -0.927 (-2.32) | 0.029 (1.93) | 1.956 (2.51) | 0.697 (6.28) | -0.879 (-5.80) | 0.053 (1.80) | - | - | - | - | - | - |
| Parking search time | -1.925 (-4.65) | 0.877 (5.41) | -1.146 (-4.74) | 0.548 (1.80) | -1.338 (-2.59) | 0.384 (6.69) | -0.756 (-2.34) | 0.135 (2.17) | - | - | - | - | - | - |
| Parking egress time | -2.246 (-1.90) | 0.990 (3.24) | -1.212 (-2.59) | 0.568 (2.00) | -1.653 (-3.31) | 0.788 (2.33) | -0.821 (-6.12) | 0.256 (4.17) | - | - | - | - | - | - |
| In-vehicle travel time* | -3.652 (-3.07) | 1.259 (2.58) | | | -4.397 (-1.84) | 1.837 (0.78) | | | -1.443 (-2.04) | 1.058 (1.92) | -3.009 (-1.81) | 1.284 (1.81) | -3.180 (-1.79) | 3.839 (3.39) |
| Trip cost* | -1.645 (-3.12) | 1.148 (4.23) | | | -2.225 (-3.86) | 1.689 (2.38) | | | -1.692 (-4.92) | 2.606 (1.83) | -1.396 (-2.72) | 1.395 (3.12) | - | - |

Notes: 1. #ASC for bus mode is set to zero for the identifiability reasons.

2. *Readers should read the parameter estimates for these variables as common for Off-street as well as On-street parking. For example, the coefficient of PV_Individualist should be read as 1.389 for *Car* in general and not solely for *Car with Off-street parking*.

3. Null loglik = -46600.68; Final loglik = -16993.45; adj. $\rho^2$= 0.627

The results of public transit attitudes are in line with those found in the literature. From the
association of four attitudes of public transit in concurrence with bus use, it is clear that safety
followed by convenience are the most detrimental factors for bus use. It can be observed that the
magnitude of mean of coefficients are higher with lower standard deviation, which shows lesser
heterogeneity in decision-making. The probability is estimated as P($\beta_{PT_Safety} < 0$) = 0.998 and
P($\beta_{PT_Convenience} < 0$) = 0.983, which indicates almost all individuals who confer more
importance on these aspects, have a lower use of bus. Further, the comfort (in negative manner)
has positive coefficient for bus indicate that people give less emphasize to it given their lower
cost affordability and lesser income (particularly who use bus in Delhi). Likewise, safety
influence attitudes towards using metro for commuting though the extent is lower compared to
bus. Convenience and flexibility are related positively with the use of metro, which is in contrary
to what observed for the bus. It shows that keeping other parameters constant, respondents are
more likely to select a metro over bus for commuting given their psychological mindset for
convenience (primarily time-reliability, please refer Table 3 and A2) and flexibility. Further it
can be recognized that respondents give least preference to walking/cycling modes looking to the
association of three LVs of PT other than safety with the NMT.

Higher and positive $\beta_{NMT_Health}$ confirm that the respondents who are health-concerned tend to
use NMT modes over all other modes given the calculated probability of this choice being close
to 1. It is also found that the more someone perceive NMT a convenient mode, the more he/she
will use NMT for short-distant commute trips. On the other hand, those who value safety and
comfort (in negative manner) feel negatively towards walking/cycling.

The built-environment factors also exhibit significant influence especially on the mode-choice
which are in line with the past studies discussed in literature. It is seen that the propensity to ride
with PT modes decreases with increase in access distance as the associated coefficients
possessed negative sign for access distance. In a similar fashion, the coefficients of egress
distance to both bus and metro are negative for the respective modes. That is, when egress
distance increases, the probability of using bus and metro for commuting decreases. It can be
noticed that the coefficients of egress distance have higher magnitude compared to those of the
access distance, which shows the respondents are more sensitive towards last-mile connectivity.
The entropy index has a negative effect on car, MTW, and NMT modes, while it is positive for
the bus and metro. It reveals that the more diversified the land-use, the higher will be use of
public modes of transport. Unlike previous works, this model shows that use of NMT decreases
for higher entropy index, though the magnitude of the coefficient is much lower. The model
results for residential density demonstrate that the use of public transport modes is positively
related to the denser urban area. Further, it can be perceived that people are likely to use MTW
or bus for travel or to walk to a greater extent when they reside in the area with higher working
population density.

Parking and travel parameters possess a considerable impact on individuals' choice behaviour.
IVTT and travel cost are negatively associated with the selection of a travel mode. Coefficient of IVTT for MTW has the highest value among the considered modes, which shows that a small rise in IVTT would have a significant effect on MTW being chosen as a travel mode. Similar effect is seen in case of metro, car, as well as NMT modes, while it is smaller for the bus. Generally, bus users are less willing to pay for improved IVTT compared to other modes (see Section 5.4), which can be considered as a potential reason for such results. The model shows that parking related attributes also have a significant influence in respondents' decision-making. Parking search time and parking egress time has negative coefficients for both the parking types and indicate that the utility decreases when the value of these variables increases. Parking duration also has a significant impact on joint mode-parking choice behaviour. It is obvious that people who need to use a parking for more duration would prefer off-street parking as captured by the model. Relatively higher magnitude of parking duration for car users (both on street and off street) shows their higher tendency to shift to off-street parking when parking duration increases. The Employer-paid parking can be treated as a very deciding factor for choosing PVs as commute modes as can be seen from the model. Here, only off-street is considered for employer-paid parking as no related on-street parking was reported in survey.

### 5.4. Willingness to pay analysis

Table 11 summarizes the distribution as well as min-max values of the estimated MWTP. Among the accessibility parameters, egress distance to metro clearly shows the largest MWTP value with less variation, indicating that commuters are more willing to pay for higher accessible metro at their workplace. This is also supported by the higher coefficient value of attitude towards PT convenience in the model. Further, it is revealed that the commuters put more value to the egress distance compared to access distance for both the PT modes. This finding is similar to one observed in meta-analysis by Ewing and Cervero (2010). Some of the respondents put negative values on MWTP, which shows that they have to be compensated by the stated amounts (for e.g., maximum of ₹17.63 per 100 meters for access distance to bus) in order to add the positivity for using bus. The derived MWTP values for IVTT-Car are found in line with the ones reported by Varghese and Jana (2018) for multitasking behaviour during commute travel using RP data (₹132.2/hr for multitasking and ₹179.2/hr for no activity), and by Athira et al. (2016) for higher income group people who commonly tend to use cars over other alternatives (₹142.19/hr), though their estimates were generalized and not alternate-specific. For IVTT-Bus, the calculated MWTP is found similar to the generalized value of travel time savings estimated by Chepuri et al. (2020) for the bus users (₹18.6/hr) in the city of Mumbai, India. In case of IVTT, car users are comparatively more willing to pay to reduce their travel time on an average, followed by MTW and metro users. A reasonable explanation for this is the income factor which plays important role in users' willingness-to-pay. Generally, rail-based transit is preferred to avoid inconvenience caused due to traffic jams on roads and hence such tendency in terms of IVTT makes sense. While Table 11 unveils the preferences among commuters for the considered attributes, these preferences are accompanied by significant taste heterogeneity, which is manifested by relatively higher standard deviations. To date, very little is known regarding this heterogeneity in context of transport mode choice. Planners and policy makers are interested in understanding the association of socioeconomic characteristics with the heterogeneity in MWTP to better plan the transportation systems. In line with this, the follow-up regression models are developed considering individual-level MWTP (conditional upon their current choice) as dependent variable and their socioeconomic characteristics (see Table 1) as the explanatory variables. Table 12 demonstrates the estimation results for the same.

Table 11 Distribution of marginal willingness to pay

| Alternative | Parameter | MWTP | | | | |
|---|---|---|---|---|---|---|
| | | Mean | SD | Median | Min | Max |
| Car | In-vehicle travel time (₹/hr) | 162.82 | 13.58 | 162.80 | 139.40 | 394.96 |
| MTW | In-vehicle travel time (₹/hr) | 64.14 | 23.78 | 57.57 | 24.15 | 178.66 |
| Bus | In-vehicle travel time (₹/hr) | 15.51 | 6.71 | 14.33 | -25.95 | 56.18 |
| | Access distance- Bus (₹/100m) | 13.72 | 6.74 | 12.26 | -17.63 | 74.51 |
| | Egress distance- Bus (₹/100m) | 24.60 | 10.61 | 22.49 | -34.59 | 83.25 |
| Metro | In-vehicle travel time (₹/hr) | 52.84 | 16.46 | 48.54 | 27.07 | 178.82 |
| | Access distance- Bus (₹/100m) | 23.78 | 16.92 | 24.09 | -27.67 | 95.53 |
| | Egress distance- Bus (₹/100m) | 36.88 | 11.88 | 34.94 | 3.94 | 120.20 |

Table 12 Marginal willingness to pay regression models

| Dependent Variable (MWTP for… ↓) | Intercept | Female | Age | Education | Income | No. of Working members | $R^2$ |
|---|---|---|---|---|---|---|---|
| Access distance- Bus (Bus users) | 0.981[a] | -0.112[b] | -0.014[b] | 0.039 | 0.022 | -0.020[b] | 0.34 |
| Egress distance- Bus (Bus users) | 1.591[a] | 0.857[b] | -0.058[c] | 0.880 | 0.052[b] | -0.030 | 0.27 |
| Access distance- Metro (Metro users) | 2.087[a] | -0.235[b] | -0.028[b] | 0.437[b] | 0.023[b] | -0.037[b] | 0.26 |
| Egress distance- Metro (Metro users) | 1.690[a] | 0.248[c] | -0.032[b] | 0.043[b] | 0.061[a] | 0.038 | 0.25 |
| In-vehicle travel time (Car users) | 8.153[a] | -0.231[a] | -0.019[b] | 1.950[b] | 0.088[a] | 1.877 | 0.36 |
| In-vehicle travel time (MTW users) | 3.838[a] | -0.739[c] | -0.007[c] | 1.309[c] | 0.085[b] | -0.943[c] | 0.21 |
| In-vehicle travel time (Bus users) | 4.766[a] | -0.551[c] | -0.057[a] | -0.367 | -0.005[c] | -0.050[b] | 0.32 |
| In-vehicle travel time (Metro users) | 3.340[a] | -0.242[a] | -0.007[c] | 0.388[a] | 0.011[a] | 0.850 | 0.29 |

Note: [a] Significant at $p < 0.01$; [b] Significant at $p < 0.05$; [c] Significant at $p < 0.10$.

Before carrying out analysis, the variables were scaled to ensure the proportionality. Though the
regression fit ($R^2$) values are relatively lower, the models are able to explain approximately 21%
to 34% of heterogeneity in MWTP. It can be seen that gender and age are significant at 90% or
higher level in all the models. It is noticed that females are less likely to pay more price in
account of the MWTP except for the egress distance for both the PT modes. The coefficient for
age is negative for all the cases which signifies that the older commuters are less willing to pay
additional amount for improvement in considered parameters. The coefficient for education level
is found significant and positive for MWTP for all but the parameters for bus. It implies that the
users are more likely to pay extra when their education level is comparatively higher, while no
significant impact of education is found for bus users. A possible reason for this is that the
commuters with higher education levels prefer to travel by modes other than bus (36% of the bus
users are non-graduates in dataset is highest among the considered modes). The coefficient for
income is negatively associated with IVTT for bus users, though the magnitude is very less. It
indicates that higher the income of bus users, lesser would be their willingness to pay for an
improved IVTT. In a reverse way, higher willingness to pay is observed with increased income
for IVTT of other modes. With the increased income, commuters tend to shift to car or metro for
more comfort (see Table 7 and 10), which could explain why they placed less importance to
MWTP for IVTT in bus. The MWTP for egress distance to metro also reveals the similar result
as the coefficient shows the positive sign. The coefficient for number of working members in
family is negatively associated with MWTP. The possible reason can be identified in an
integration with the household structure. For example, MWTP for a person from household
having five people with two working members could be lesser than that for a person from
household having two people with both as working members, given that both households have
nearly same income. Similar to the travel parameters, MWTP for reducing parking search time
and egress time to destination can also be analysed considering the parking price coefficient
instead of travel cost coefficient. But it wasn't possible for this case as the current parking price
structures in the study location are equivalent for both types of parking.

**6. Discussion and Conclusions**

This paper investigates commuters' travel behaviour by considering parking choice as an
endogenous decision within travel mode choice framework. Using the RP data collected through
individual survey, an HDCM with hierarchical Bayes estimator was formulated to examine the
relationships between commuters' subjective evaluations as well as objective characteristics and
their choice behaviour. As the proposed model has an ability to provide individual-specific
parameter estimation through simulation, it is possible to understand the heterogeneity in
respondents' willingness-to-pay.

Model estimation results show that various psychological factors attributable to the travel modes
and parking types are significant in determining commuters' choice behaviour. The model
showed that the attitudes towards safety, comfort and convenience are strong drivers for respondents' parking choice as well as mode-shift decisions in general. Particularly for PVs, individualist attitude is seen to have significant influence on both car and MTW choices. In countries like India, vehicle ownership is seen as a reflection of the wealth and social status. This in turn encourages individuals to use their private vehicles due to lack of ample knowledge of sustainability. Further, there is no policy in place to control the vehicle ownership. Even the parking charges are very low compared to other similar size cities across the world (MoHUA, 2019). City authorities should focus on these aspects as they are key elements in habit-breaking (of PV-use). Apart from improving the public transit service quality, educational program should be organized which can strengthen individuals' perception and responsibility towards better environment. As CBD areas remain congested especially during the peak-hours in Delhi (and in India), revision of on-and off-street parking pricing and provision of congestion pricing scheme should be considered in policy development which can significantly limit the PV-use. It can also increase the use of transit in inner city areas which are generally well connected by bus and metro services.

Public transit users put highest importance to safety and convenience; and hence, it should be of utmost importance to planning authorities also to promote the PT use. Provisions for adequate police patrolling in nearby areas of PT stations with ample lighting and security cameras will positively affect the perceptions towards security for commuters especially who travel in dark. As women are more concerned about safety in PT and NMT (Table 7 & 8), provisions such as security presence on roads and in vehicles, regular safety audits in sensitive areas, cameras and GPS systems, stop-on-request, and emergency buttons in vehicles should be considered for improved safety. The lower-income group tend to live farther from the city center and CBDs. They don't have much viable transport options but to use MTW for commute because of non-affordability of car and poor first/last mile connectivity issues with transit, which is also confirmed by this study. These can be the primary reason for higher MTW ownership in developing countries. The choice model shows that rise in access/egress distance make transit modes less attractive. Willingness-to-pay analysis shows the significance of access/egress characteristics on selection of main transport mode. Hence, due consideration should be given to improve last-mile connectivity of public transport modes. Increasing coverage of public transit seem to be a solution to address this issue. Practically, it is difficult to provide fixed-guideway mass transit (like metro) in all areas of the city due to higher marginal costs. Hence, such area should be facilitated by lower category of transit service like bus, regulated para-transit services. Apart from this, providing efficient feeder services can confine the existing transit users escaping from using the service and attract private vehicle users towards transit (Saiyad et al., 2021). Provisions for ample parking space near important transit terminals can also increase the PT ridership in terms of main-haul trip. In addition, integration of PT system (bus + metro in this case) in terms of schedule, fare, and stops seem to increase the PT ridership as it can effectively compete with private vehicles usage (Zimmerman and Fang, 2015). Such integration with common mobility fare card and well-organized feeder-service may prove to be a promising action for promoting PT.

The model results illustrate key findings for NMT-use: all four latent variables – health, safety, comfort and convenience – have strong impacts on peoples' intention to use NMT. In most Indian cities including Delhi, a common problem is observed that the sidewalks are generally occupied by the vendors and in poor condition at many locations forcing pedestrians to get down on the carriageway. Also, bicyclists generally ride in mixed traffic due to absence of dedicated path or hindrance by vendors (if present) which declines bicycle-usage significantly as a result of lesser safety and convenience. Policy interventions have to be focused on improving peoples' satisfaction which may in turn improve the NMT share. The educational program focusing on coexistence of transport modes, prioritization of NMT users on roads, and benefits of NMT-use on health and environment might be accompanied with these interventions for better outcomes. The *Street Vending Policy* should appropriately address the vending management in cities to overcome the stated issues. The guidelines published in Tender S.U.R.E. (Specifications for Urban Road Execution) and implemented in the city of Bangalore can be considered in this line (Ramanathan, 2011).

This study also revealed important results regarding the shift between parking types. Authors have tried to increase the search time and egress time by 50% for parking to assess how the shift takes place. Table 13 summarizes the current market share, model-predicted market share, and change in share with respect to increase in parking parameters. It clearly reveals that major shift occurs between parking types instead of switching travel mode when parking attributes of one of the types rises. Also, the car users seem more vulnerable to these changes compared to MTW users. Individuals of current off-street parking users shift more towards metro while individuals of current on-street parking users shift more towards bus. Since bus is more accessible than metro, these findings reveal that current on-street parking users are less prone to walk more to get transit even if they shift from PV to PT mode. The last two rows of the Table 13 try to identify the users' response to the increase in parking attributes for both parking-types. It reports the interesting results that significant shift would take place between PV modes only (i.e., from car to MTW) which reveals that these respondents are highly habitual to use PVs. The basic reasons for such behaviour are latent variables which are discussed earlier. Even though this study couldn't identify the monetary value of parking search time and egress time, these results can be used to compare the valuation of search time and egress time. On-street parkers put more value to egress time while off-street parkers to the search time. From the combined scenarios, overall valuation is higher for search time than that for egress time. Lastly, very less shift is seen towards NMT modes except when parking egress time increases. These group of respondents may not be willing to spend much time to search for vacant parking place and then walk to the destination as they travel for much lesser time between origin and destination.

Table 13 Change in market share (%) under different parking scenarios

| Scenario | Car_Off | Car_On | MTW_Off | MTW_On | Bus | Metro | NMT |
|---|---|---|---|---|---|---|---|
| Current market share | 17.28 | 7.64 | 11.30 | 13.95 | 24.58 | 16.61 | 8.64 |
| Predicted market share | 17.65 | 6.98 | 10.47 | 14.78 | 25.25 | 16.28 | 8.60 |
| Increase off-street parking search time | -2.55 | 1.98 | -2.17 | 1.69 | 0.66 | 0.77 | 0.00 |
| Increase on-street parking search time | 1.24 | -2.17 | 0.90 | -1.73 | 0.49 | 0.39 | 0.06 |
| Increase on-street parking egress time | 1.64 | -2.43 | 1.08 | -1.97 | 0.57 | 0.48 | 0.07 |
| Increase off-street parking egress time | -3.15 | 2.67 | -2.32 | 1.75 | 0.64 | 0.89 | 0.15 |
| Increase parking egress time for both types | -2.35 | -1.05 | 0.44 | 1.10 | 0.62 | 0.68 | 0.19 |
| Increase parking search time for both types | -1.75 | -0.93 | 0.44 | 0.97 | 0.48 | 0.91 | 0.10 |

Another important observation from the present study is the positive association of employer-paid parking with PV-use. It can create hindrance in attaining sustainability if there are no counter policies like parking cash-out or mass transit subsidies established alongside. The related concerns are well discussed by Shoup (2005) and Parmar et al. (2021). As a policy measure, urban local bodies may introduce *parking credit system* for encouraging PV users to utilize PT system for select work trips of the week. For such trips, the PV users can earn parking credit points which may be encashed against parking charges at sites of higher parking cost (e.g., shopping areas). This may gradually improve the PT modal share for commute trips. Moreover, appropriate parking fare structure including differential parking pricing for on-street and off-street lots should be formulated to effectively manage the parking supply as well as to confront the serious on-street parking issues. Though there is a less room for intentional changes in parking search time and egress time compared to parking pricing, building remote parking lots in the peripheral parts of CBDs with proper integration with transit in inner parts can be considered as a long-term sustainable solution.

In general, this study provides important evidence on how commuters' attitudes and built-environment characteristics, in addition to the alternate-specific attributes, can act as decisive factors in determining their travel mode and parking-type choice simultaneously. Based on the choice model, marginal willingness-to-pay has been determined for the selected parameters. Planners and policymakers can thus get better insights on the mode choice behaviour as well as possible taste heterogeneity in WTP. This is crucial for planners and PT service providers in understanding the policy implications and making better development strategies. From methodological point of view, this paper develops Bayesian HDCM model using simultaneous approach which minimize the overall error in the model and yields more realistic econometric model. When compared with the maximum simulated likelihood, especially with higher number of latent variables, Bayesian estimation clearly outperforms this classical estimation in terms of the computational efficiency.

It should be noted that the model results and MWTP values could not be directly compared to developed countries just like rapidly urbanizing developing countries in south Asia as income and population densities play major roles in this case. To overcome the limitations related to parking cost in this study, a combined framework of RP-SP survey data can be deployed to assess commuters' response to various parking pricing policies in addition to other attributes. This framework can also assist in understanding commuters' willingness-to-pay for various parking attributes explicitly.

1 **APPENDIX**

2 Table A1 Indicators for latent variables for private vehicles

| Latent Variable | Indicator |
|---|---|
| Individualist | Individualist1. I have social status. I will use my vehicle only. |
| | Individualist2. I can afford PV and related costs. Why should I go for PT!! |
| | Individualist3. Driving is fun and relaxing. |
| | Individualist4. Driving to destination provides safety and privacy compared to other alternatives. |
| Pro-environment | Pro-environment1. Private vehicles are the reason for congestion and pollution. |
| | Pro-environment2. I can contribute to make difference on environmental problems. |
| | Pro-environment3. People should be made aware of environmental concerns. |
| Economy | Economy1. Higher tax should be levied from PV users against congestion. |
| | Economy2. The fuel price should be increased to limit PV-use. |
| | Economy3. Parking fees should be high to limit PV-use. |
| Comfort | Comfort1. Crowding and comfort are main reasons why I prefer PV over PT. |
| | Comfort2. Using PV is time saving and reliable. |
| Flexibility | Flexibility1. PV offers flexibility of choosing route and I can make multiple stops if required. |
| | Flexibility2. It is difficult to find parking space nowadays. |
| | Flexibility3. Looking to congestion, I worry to be on time at my destination while using my car. |

4 Table A2 Indicators for latent variables for public transport

| Latent Variable | Indicator |
|---|---|
| Comfort | Comfort1. Crowding and comfort are main reasons why I prefer PV over PT. |
| | Comfort2. It is annoying to take multiple stops/transfers while travelling by PT. |
| | Comfort3. Waiting for bus is annoying. |
| | Comfort4. Taking PT is difficult when I travel with bags/luggage. |
| Convenience | Convenience1. We need more accessible PT to limit PV-use. |
| | Convenience2. I am not sure whether I will be on time to my destination while traveling by Bus. |
| | Convenience3. I will be on time if travel by Metro. |
| Safety | Safety1. Lesser chances of accidents if we use PT. |
| | Safety2. PT (including stops) is not safe from theft when comparing with other alternatives. |
| | Safety3. I do not like to be surrounded by unknown people. |
| Flexibility | Flexibility1. Services (like shops and food stalls) make waiting more pleasant. |
| | Flexibility2. Getting a seat from time-to-time is not my concern. |
| | Flexibility3. Waiting is not much annoying as I can find schedules from internet. |

2 Table A3 Indicators for latent variables for non-motorized modes

| Latent Variable | Indicator |
|---|---|
| Health | Health1. It is healthier to use walk/bicycle mode. |
| | Health2. I will prefer walk/bicycle looking to the environmental issues. |
| | Health3. Cycling is relaxing and fun. |
| Safety | Safety1. Lack in infrastructure makes these modes most vulnerable. |
| | Safety2. Walking is not safe without CCTV surveillance. |
| | Safety3. It is not preferable during late evening and night. |
| Comfort | Comfort1. It is annoying during summer and rains. |
| | Comfort2. No safe parking for bicycle, issues of theft. |
| Convenience | Convenience1. Walking/Cycling are flexible and time-saving for short trips. |
| | Convenience2. Almost no travel cost for these modes. |

4 Table A4 Indicators for latent variables for parking types

| Latent Variable | Indicator |
|---|---|
| Safety | Safety1. My vehicle will not be safe if parked on-street for long time. |
| | Safety2. On-street parking is hazardous for traffic. |
| Convenience | Convenience1. I hate walking. On-street parking is superior for me. |
| | Convenience2. Getting on-street parking space is difficult and time-consuming. |
| | Convenience3. Compared to cruising for on-street parking, I will prefer to walk from off-street parking. |